\documentclass{PoS}
\usepackage{amsmath}
\usepackage{slashed}
\usepackage{soul}
\newcommand{\matel}[3]{\langle #1 \mid\! #2 \mid\! #3 \rangle}

\title{Resonance study of SU(2) model with 2 fundamental flavours of fermions}

\ShortTitle{Resonance study of SU(2) model}

\author{\speaker{Tadeusz Janowski}\\
School of Physics and Astronomy, 
University of Edinburgh,\\
James Clerk Maxwell Building,
Peter Guthrie Tait Road,
Edinburgh,
EH9 3FD\\
        E-mail: \email{t.janowski@ed.ac.uk}}

\author{Vincent Drach\\
        School of Engineering, Computing and Mathematics, University of Plymouth\\
        2 Kirkby Place, Drake Circus, PL4 8AA Plymouth\\
        E-mail: \email{vincent.drach@plymouth.ac.uk}}

\author{Sasa Prelovsek\\
        Faculty of Mathematics and Physics, University of Ljubljana, Slovenia\\
        Jozef Stefan Institute, 1000 Ljubljana, Slovenia\\
        Institut f\"ur Theoretische Physik, Universit\"at Regensburg,\\ D-93040 Regensburg, Germany.\\
        E-mail: \email{sasa.prelovsek@ijs.si}}

\abstract{Composite Higgs models are promising candidate models to address
the long-standing naturalness problem in the Standard Model.
Among them, the most minimal one is the SU(2) with 2 flavours of
fermions in the fundamental representation of the gauge group.
An important prediction in these models is the existence of resonance
spectrum in vector boson scattering. Here we study the lowest such resonance, which is the equivalent of rho
resonance in QCD. We describe the scan of the parameter space using the
clover-improved Wilson fermions with Symanzik improved gauge action and
then show the first results for the mass and width of the rho resonance
in this model.}

\FullConference{37th International Symposium on Lattice Field Theory - Lattice2019\\
		16-22 June 2019\\
		Wuhan, China}

\begin{document}

\section{Introduction}

With the discovery of the Higgs boson, the naturalness problem is confirmed to be a pressing mystery of the Standard Model (SM).
It is known that the observed Higgs mass of around 126 GeV arises as a result of cancellation between the bare Higgs mass parameter in the SM Lagrangian and the radiative corrections. The latter are expected to be of the order of the cutoff of the theory and if we assume that the Standard Model is valid up to the Planck scale O($10^{19}$ GeV) this translates to a fine tuning of 1 part in $10^{17}$.

Several solutions to this problem have been proposed in the literature. 
These include supersymmetry, extra dimension models and composite Higgs models. In the latter, the Higgs boson is treated as a composite state made out of `techniquarks' or `hyperquarks' held together by a new type of strong interaction. The Higgs mass is then naturally connected to the compositeness scale of the new strongly-interacting theory.

We focus on one such model which is the minimal extension of SM that has all the desired properties of the composite Higgs model. We perform lattice simulation of  this model in isolation, i.e. decoupled from the   SM sector. It features stable particles that can not decay via the new strong interaction, as well as resonances. Our aim is to determine the mass as well as the width of the lightest vector resonance by simulating the scattering on the lattice. 

\section{The model}

We consider SU(2) gauge theory with 2 fundamental flavours of fermions~\cite{Cacciapaglia:2014uja}
\begin{equation}
\label{model}
    \mathcal L = -\frac{1}{2} Tr \left( F^{\mu\nu}F_{\mu\nu} \right) + i \bar u \slashed D u + i \bar d \slashed D d.
\end{equation}
While superficially similar to two-flavour QCD, as a consequence of equivalence of fundamental and anti-fundamental representation, this model has an enhanced flavour symmetry, i.e. $SU(4)$, which connects the left-handed Weyl fields with charge-conjugated right-handed Weyl fields.
The formation of the condensate $\Sigma = \langle \Psi_i \Psi_j^T \rangle$ where $\Psi^T = (u_L,\: d_L,\: -(i\sigma^2_c)(i \sigma^2_s) u_R^* ,\: -(i\sigma^2_c)(i \sigma^2_c)d_R^*)$ breaks the SU(4) flavour symmetry of $\Psi$ into a subgroup which preserves the condensate $U \Sigma U^T = \Sigma$. Hence the symmetry breaking pattern is $SU(4)\to Sp(4)$ or equivalently $SO(6) \to SO(5)$, which is the next-to-minimal composite Higgs model. This generates 5 Goldstone bosons, which we will refer to as `pions'.

Although different condensates correspond to different physics in the model including electroweak interactions, they are all equivalent for the model in isolation which we study here (\ref{model}).
This is important for the lattice calculation where we are forced to add a fermion mass term in (\ref{lat_action}) below, which explicitly breaks SU(4) symmetry to a particular Sp(4) subgroup.  

Irrespective of the actual symmetry-breaking scenario, three of the Goldstone bosons will eventually become the longitudinal components of the W and Z vector bosons. 
This has the effect that, due to Goldstone boson equivalence theorem, any dynamics related to the Goldstone bosons should be visible in vector boson scattering at high energies.
Studying the dynamics of pion scattering in these models gives us therefore direct access to the phenomenology of vector boson scattering, allowing us to predict widths and masses of resonances which may appear in the experiments in the future.

The pions are in the {\bf 5} representation of SO(5), meaning that the 2-pion system can be in $\mathbf 5 \otimes \mathbf 5 = \mathbf{14} \oplus \mathbf{10} \oplus \mathbf 1 $.
The {\bf 14} representation is non-resonant, because the lightest bound state with this flavour quantum number has to consist of four hyperquarks. 
The scattering length in the {\bf 14} representation has been studied in \cite{Arthur:2014zda}.
The {\bf 10} has the same quantum numbers as a vector resonance, the analogue of $\rho \to \pi\pi$ in QCD and it is the main subject of study here.

From the phenomenological standpoint, if such a vector resonance were too light or too narrow it would be visible in the vector boson scattering cross-section. This can be used to produce exclusion plots like was done for the Minimal Walking Technicolor in Fig. 10 of \cite{Aad:2014cka}.
The fact that this is not the case imposes experimental constraints on the model. These constraints can be sharpened by the first-principle analysis of the $\rho$ resonance in this model from Lattice QCD, which is the subject of the remainder of this proceeding.

\section{Lattice setup}

To study strongly-coupled model (\ref{model}), 
we simulate lattice action:
\begin{align}
\label{lat_action}
    S &= \frac{\beta}{2} \sum_{x, \mu, \nu} c_0 \mathrm{Re Tr} P_{\mu \nu}(x) + c_1 \mathrm{Re Tr} \left( R_{\mu\nu}(x) + R_{\nu\mu}(x) \right)\\
      &+ \sum_{x,\mu} \bar \psi(x) \left( m_0 + 4 \right)\psi(x) - \frac{1}{2} \bar \psi(x+\mu) U_\mu (1 - \gamma^\mu ) \psi(x)  ~+~ \frac{c_{sw}}{2} \sum_{x, \mu<\nu} \bar \psi(x) \sigma_{\mu\nu} \hat F^{\mu\nu} \psi(x),\nonumber
\end{align}
where $P_{\mu\nu}$ is the plaquette, $R_{\mu\nu}$ is a 2x1 rectangular loop, $\hat F_{\mu\nu}$ is the usual Wilson clover term and $\beta=4/g_s^2$ is related to the gauge coupling $g_s$.
In other words, we use a clover-improved Wilson action and Symanzik improved gauge action with the following parameters: $c_{sw} = 1$, $c_0 = 5/3$ and $c_0 + 8 c_1 = 1$. 
The presence of bare mass term and the Wilson term explicitly breaks the SU(4) flavour symmetry to an Sp(4) subgroup. Since we are ultimately interested in a theory with massless fermions we will have to take the limit of vanishing renormalised fermion mass, or equivalently vanishing pion mass.

\begin{figure}[]
    \begin{center}
        \includegraphics[width=0.55\linewidth]{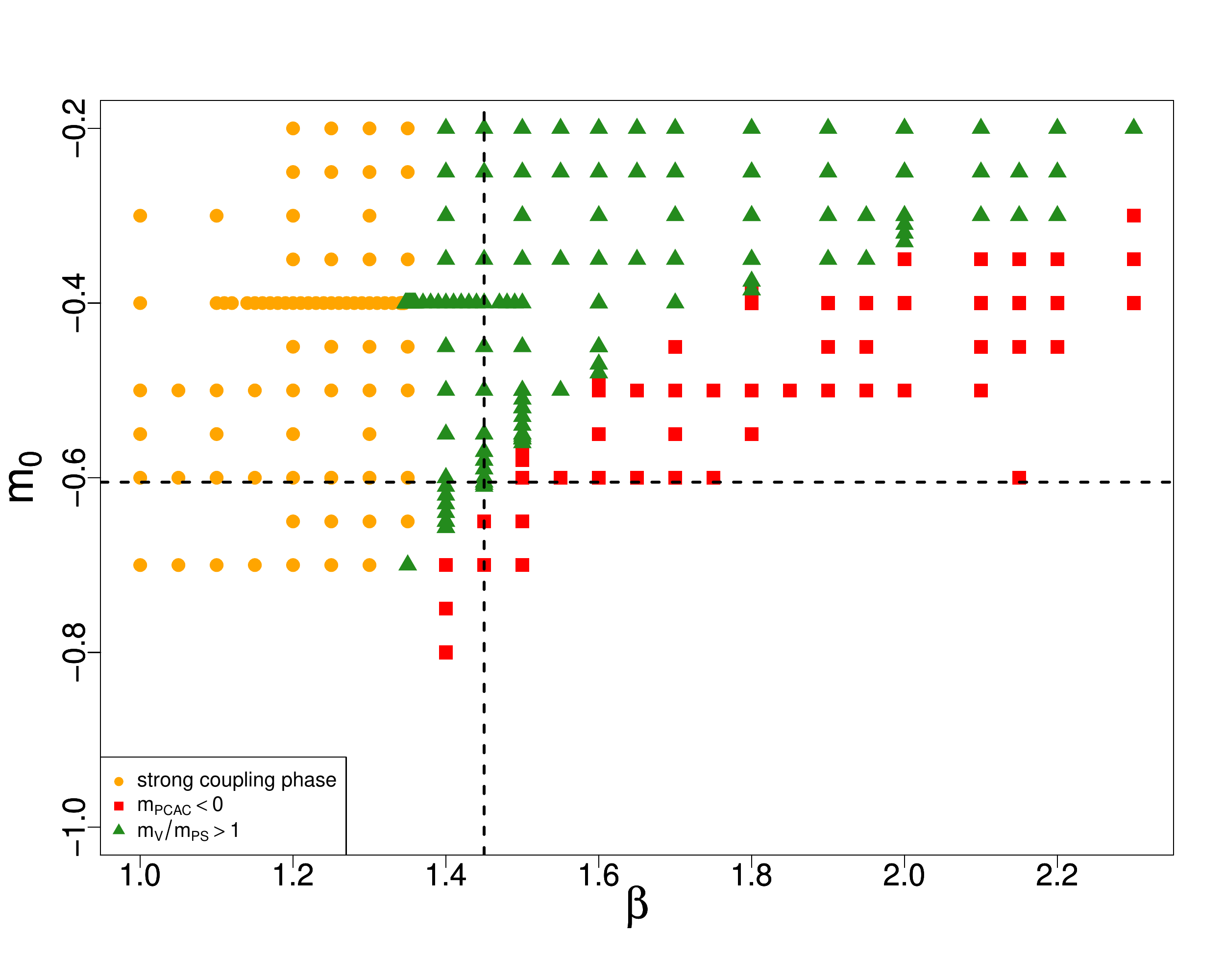}
    \end{center}
    \caption{Phase diagram of the SU(2) model (\ref{lat_action}) with 2 flavours of fermions in the fundamental representation with Symanzik improved gauge action and $c_{sw} = 1$: $m_0$ and $\beta$ are related to bare fermion mass and  gauge coupling, respectively. }
    \label{fig:phase}
\end{figure}

We have scanned the parameter space of the model for various bare fermion masses ($m_0$) and values of $\beta$. This is summarised in figure \ref{fig:phase}.

To study scattering we need a point which is in the physical phase with the renormalised fermion mass $m_{PCAC} > 0$ and on the right side of the bulk phase transition - these are denoted by the green points of the plot.
In addition, we require that $m_\rho > 2 m_\pi$ for the $\rho$ meson to be a resonance and not a stable state.
We have found one point in the phase space diagram which satisfies the above constraints:
$\beta = 1.45$ with $m_0 = -0.6050$. 
With these parameters we obtain $a m_\pi = 0.20213(6)$ on the $24^3$ and $a m_\pi = 0.22467(11)$ on the $16^3$ ensemble.
and $a m_\rho^{naive} = 0.444(9)$, where $m_\rho^{naive}$ is the effective mass extracted from the vector two-point function.
We have generated two volumes with these parameters: $16^3 \times 32$ and $24^3\times 48$. We have produced 1354 and 2551 trajectories respectively.  When calculating the propagators we use the `periodic + antiperiodic' (P+A) boundary conditions in the time direction. The purpose of this is to increase the effective time extent by a factor of 2, which reduces the contributions of finite temperature (or `around-the-world') effects.


To calculate the resonance parameters we follow  the methods used extensively in  QCD studies of $\rho\to\pi\pi$. We construct a matrix of correlation functions $C_{ij}(t) = \matel{0}{O_i^\dagger(t)O_j(0)}{0}$ where the operators $O_i$ have the same quantum numbers as the $\rho$ resonance, i.e. transform under the {\bf 10} representation of the flavour group and have a non-zero overlap with the angular momentum $j=1$ state.
The specific operators we use are:
  \begin{align}
  \label{O}
    O_1(t) &=O^{\pi( \mathbf p)\pi(0)}= \left(\sum_{x} \bar \psi(x) \gamma^5 \psi(x) e^{i \mathbf p \cdot \mathbf x} \right) \left( \sum_y \bar \psi(y)\gamma^5 \psi(y) e^{i \mathbf y \cdot \mathbf 0}\right),\\
    O_2(t) &=O^{\rho}= \sum_x \bar \psi(x) (\gamma \cdot \hat {\mathbf p})  \psi(x) e^{i \mathbf p \cdot \mathbf x},\nonumber
  \end{align}
  where $\psi$ denote the light fermion fields,  $\mathbf p=\mathbf P$ is the 3-momentum and $\hat{\mathbf p} = (0, \mathbf p/|\mathbf p|)$.
  In this project we focus on two values of total momenta: $\mathbf P = (0,0,1)$ and $\mathbf P = (1,1,0)$.
  
 Wick contractions for the correlators with operators (\ref{O})  lead to the following diagrams:
	\begin{align}
	  C_{11}(t) &= \vcenter{\hbox{\includegraphics[width=0.13\textwidth]{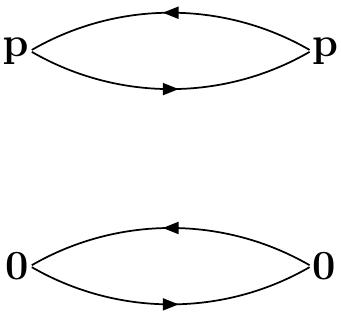}}}-\vcenter{\hbox{\includegraphics[width=0.13\textwidth]{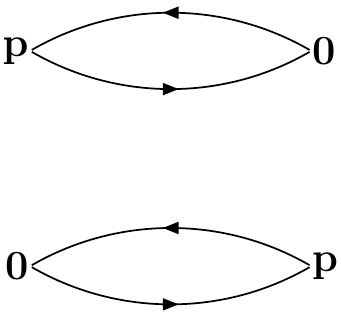}}}+\vcenter{\hbox{\includegraphics[width=0.13\textwidth]{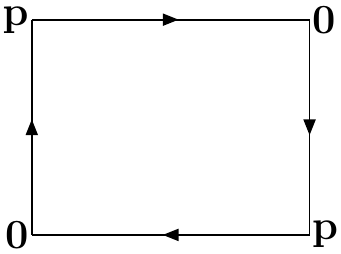}}}\nonumber\\
	  &+\vcenter{\hbox{\includegraphics[width=0.13\textwidth]{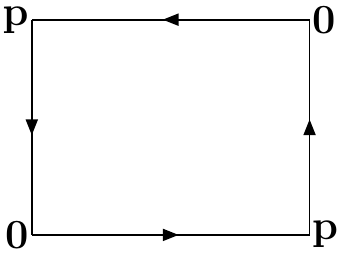}}} - \vcenter{\hbox{ \includegraphics[width=0.13\textwidth]{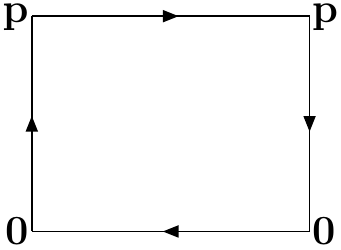}}} 
	  - \vcenter{\hbox{\includegraphics[width=0.13\textwidth]{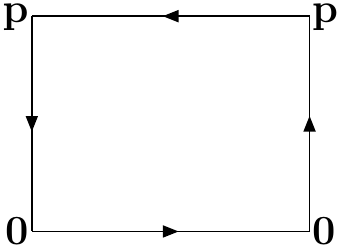}  }}
	\end{align}
	\begin{align}
	  C_{12}(t) &=-C_{21}^*(t)=\vcenter{\hbox{\includegraphics[width=0.13\textwidth]{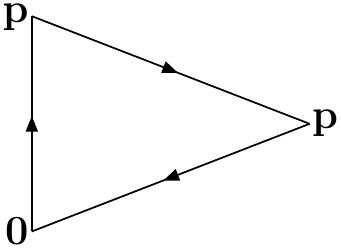}  }} - \vcenter{\hbox{\includegraphics[width=0.13\textwidth]{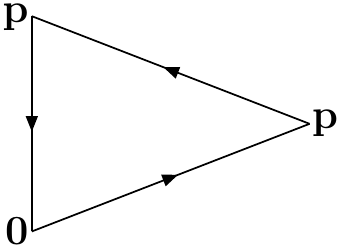}  }}
	\end{align}
	\begin{align}
	  C_{22}(t) &=\vcenter{\hbox{\includegraphics[width=0.13\textwidth]{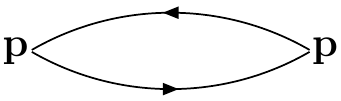}  }}
	\end{align}
    These  contractions  require all-to-all propagators and we employ technique with U(1) stochastic sources from \cite{Aoki:2007rd} that satisfy  $\tfrac{1}{N_R}\sum_{j=1}^{N_{R}} \xi^{j \dagger} (\mathbf x) \xi^j(\mathbf y) \xrightarrow{N_{R} \to \infty} \delta(\mathbf x - \mathbf y)$. We use $N_R=3$, which is sufficient to reduce the stochastic noise.
  
\begin{figure}
        \begin{center}
            \includegraphics[width=0.5\textwidth]{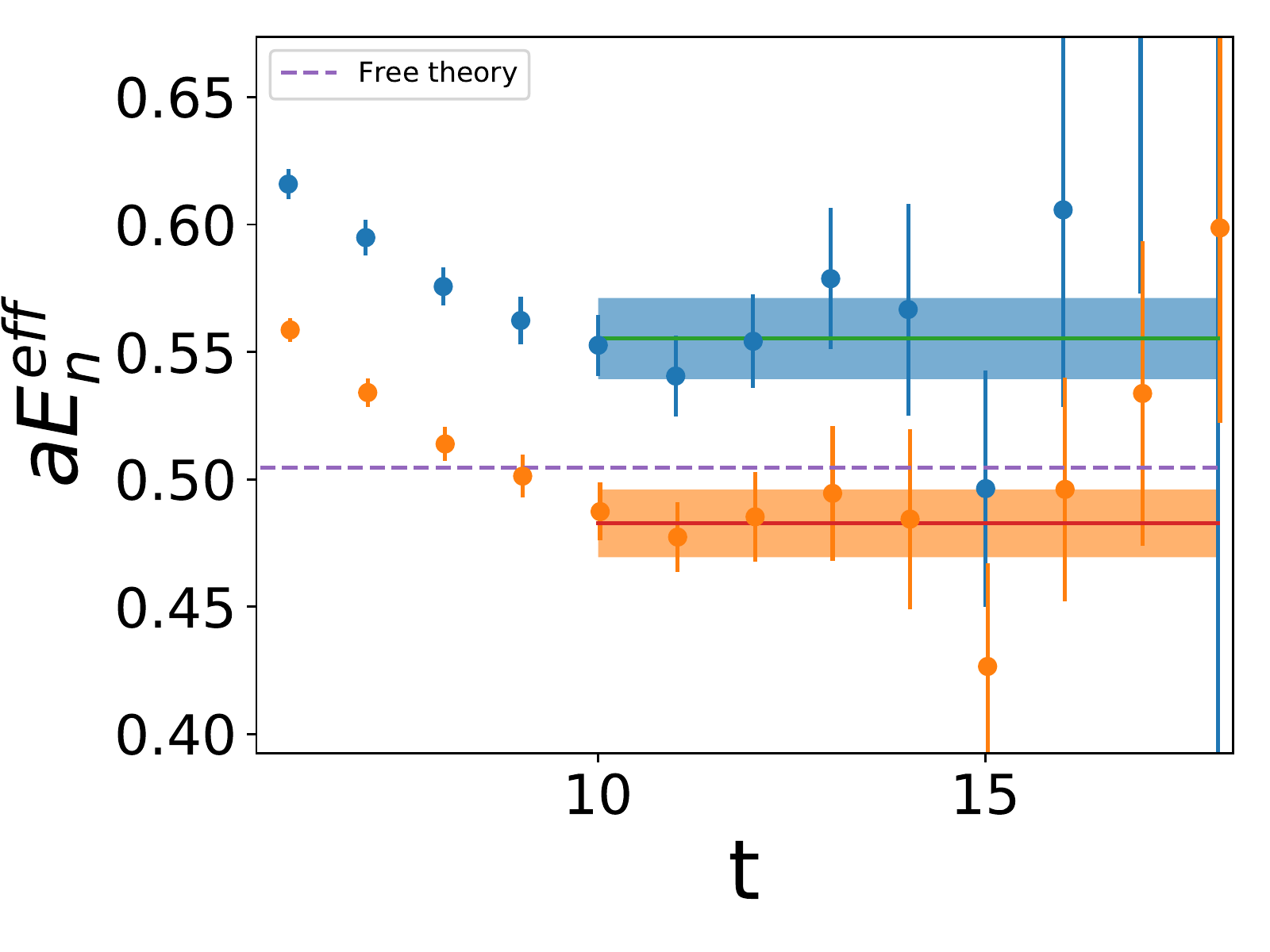}
        \end{center}
        \caption{Effective energies associated with different eigenvalues of the generalised eigenvalue problem for $N_L=24$ and $\mathbf P=(0,0,1)$. Dashed line   indicates non-interacting $\pi\pi$ energy.} 
        \label{fig:effe}
    \end{figure}

 Resulting correlation functions lead to  eigen-energies $E_n$ of the two-pion states based on 
  \begin{align}
    C_{ij}(t) \equiv \matel{0}{O_i^\dagger(t) O_j(0)}{0} &= \sum_{n} \matel{0}{O_i^\dagger}{n}  ~e^{-E_nt}~
   ~ \matel{n}{O_j}{0} ~. 
     \end{align}
     Widely used GEVP  method is employed for that, where eigenvalues   of   $C(t)u^n(t)=\lambda_n(t)C(t_0)u^n(t)$ render eigen-energies via $\lambda_n(t)\xrightarrow{large ~t }Ae^{-E_n (t-t_0)}$.   In our analysis we choose $t_0 = 4$ and verify agreement for $t_0=3-5$.  The plateaus in effective energy  
  \begin{equation}
      E^{eff}_n(t) = \log(\lambda_n(t)/\lambda_n(t+1))
  \end{equation}
  are related to energies $E_{n=1,2}$ of eigenstates. 
  As an example, we show the effective energies on the $24^3$ ensemble for $\mathbf P=(0,0,1)$ in Fig. \ref{fig:effe}.
  We observe clear plateaus for both energy levels.

    Extracted energies of two-pions in the finite volume are related to infinite-volume scattering amplitudes $S(E)=e^{2i\delta(E)}$  via rigorous L\"uscher's formalism~\cite{Luscher:1990ux, Rummukainen:1995vs}.  
    In moving frames and irreducible representations corresponding to our operators, the relations between  $\delta$  and $E_n$ are given in the table below~\cite{Feng:2010es}~(\footnote{Sign in front of $Z_{22}- Z_{2(-2)}$ is opposite since  \cite{Feng:2010es} uses $Y_{lm}^*$ instead of $Y_{lm}$ in definition of $Z_{lm}$. })  :
   \begin{center}
  \begin{tabular}{c|c|c|c}
      $\mathbf P$ & group &repr & $\tan \delta_1$\\
    \hline
$(0,0,1)$ & $D_{4h}$ & $A_2^-$ & $\frac{\pi^{3/2}q\gamma}{Z_{00}(1;q^2) + \frac{2}{\sqrt 5 q^2} Z_{20}(1;q^2)}$\\
$(1,1,0)$ & $D_{2h}$ & $B_1^-$& $	\frac{\pi^{3/2}q\gamma}{Z_{00}(1;q^2) - \frac{1}{\sqrt 5 q^2} Z_{20}(1;q^2) -i\frac{\sqrt{3}}{\sqrt{10}q^2}\left( Z_{22}(1;q^2) - Z_{2(-2)}(1;q^2) \right)}$
  \end{tabular}
  \end{center}
Here  $q = \frac{2\pi}{L} \sqrt{E_{CM}^2/4 - m_\pi^2}$ and      $ Z_{lm}(s,q^2) = \sum_{n\in P_d}Y_{lm}(n)/(q^2-n^2)^s$ \cite{Feng:2010es}. 
Every energy level $E_n$  renders certain $E_{CM}$ and   phase shift.

\begin{figure}
      \begin{center}
          \includegraphics[width=0.8\textwidth]{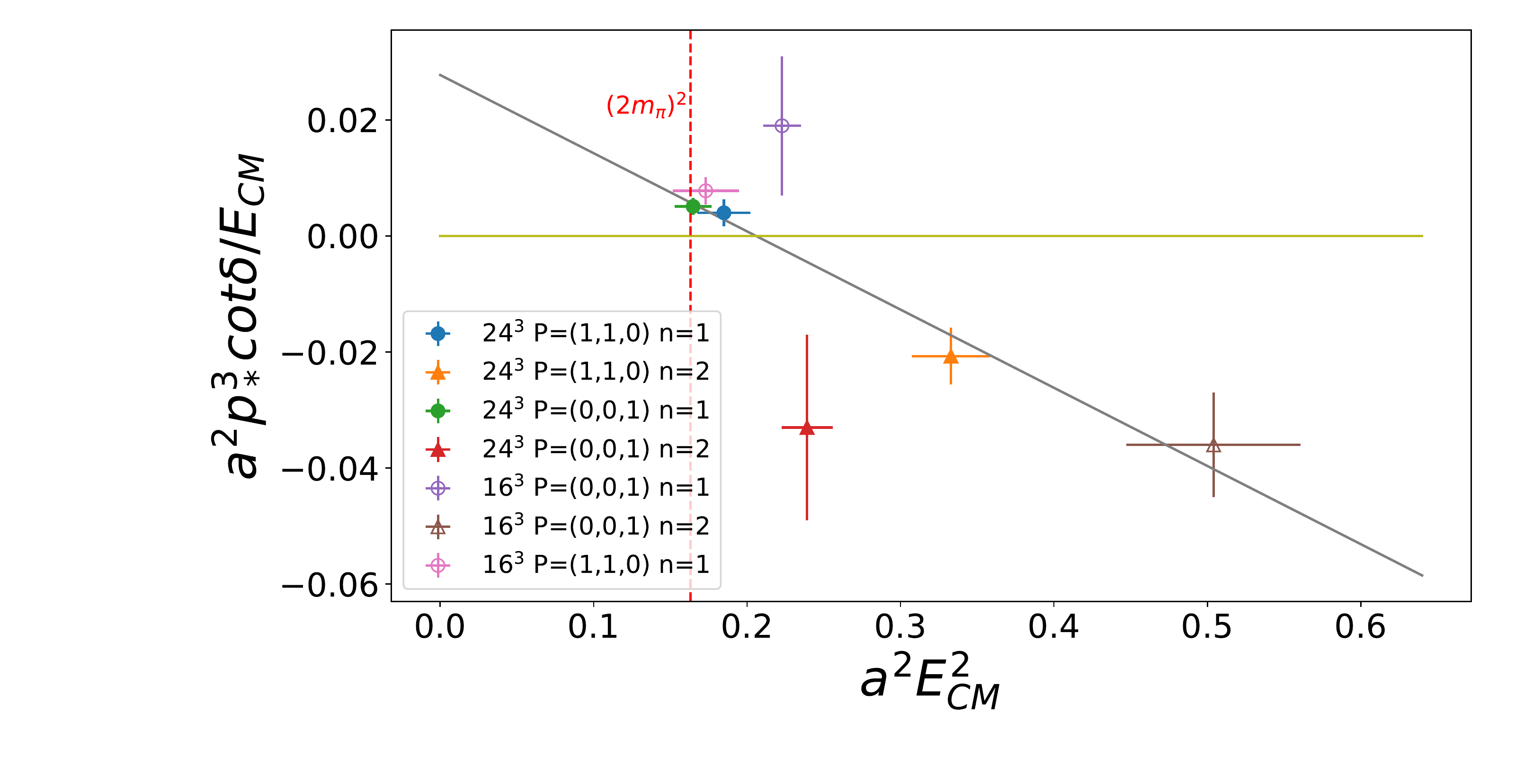}
          \caption{The plot of $a^2 p^3_* \cot \delta/E_{CM}$ as a function of the squared centre-of-mass energy. Points with error bars correspond to energy levels from different ensembles and/or total momenta $\mathbf P$. Linear dependence (\ref{linear_fit}) is expected for Breit-Wigner resonance: $g_{\rho\pi\pi}$ can be read off from the slope and  $M_\rho^2$ from the x-intercept. 
          }
          \label{fig:mainplot}
      \end{center}
  \end{figure}

The energy-dependence of the phase shift is expected to have a resonance form in the vicinity of a vector resonance and we assume the Breit-Wigner form. 
  \begin{equation}
    \tan \delta_1(E_{CM}) =  \frac{E_{CM}~\Gamma_\rho(E_{CM})}{M_\rho^2-E_{CM}^2}, \quad p_*=\sqrt{\frac{E_{CM}^2}{4}-m_\pi^2}~,\quad \Gamma_\rho(E_{CM})= \frac{g_{\rho\pi\pi}^2}{6\pi}\frac{p_*^3}{E_{CM}^2}
  \end{equation}
  The resonance width $\Gamma_\rho$ is parametrized in terms of the coupling $g_{\rho\pi\pi}$  and 
  strongly depends on the phase space. We extract the resonance mass $M_\rho$ and the coupling $g_{\rho\pi\pi}$  by fitting the quantity
  \begin{equation}
  \label{linear_fit}
      \frac{p_*^3 \cot \delta}{E_{CM}} = \frac{6 \pi}{g_{\rho\pi\pi}^2} \left( M_\rho^2 - E_{CM}^2 \right)
  \end{equation}
 that is shown in Fig. \ref{fig:mainplot}.
  We note that $p_*^3 \cot \delta/E_{CM} $ is linear as a function of $E_{CM}^2$  and the desired parameters, $g_{\rho\pi\pi}$ and $M_\rho^2$, can be read off from the slope and the x-intercept respectively.

  Fit of the phase shifts (\ref{linear_fit}) in Fig. \ref{fig:mainplot} gives
   \begin{equation}
  \label{result_res}
   g_{\rho\pi\pi} = 11.6 \pm 1.8 ~,\quad aM_\rho = 0.458 \pm 0.067~.
 \end{equation}
  This is in agreement with the `naive' $\rho$ mass extracted from the vector two-point function of $a\mathrm{m_\rho^{naive}=0.444\pm 0.009}$ since the rho resonance lies just above the threshold and is therefore narrow.
 To investigate the potential influence of the finite volume effects on the $16^3$ ensemble ($m_\pi L = 3.5$), we have also tried fitting the $24^3$ ensemble only. This  renders  $g_{\rho\pi\pi}^{(24)} = 10.7\pm 2.3$ and $a M_\rho^{(24)} = 0.445\pm 0.095$, which is compatible with our main result (\ref{result_res}).  
We note that  the coupling has value of  $ g_{\rho\pi\pi}^{QCD} \simeq 6$ in QCD.
  
 This is the first result for the resonance mass as well as the width in composite Higgs models to date. The above results apply to a certain value of  a (small) fermion mass and the lattice spacing. In order to obtain the `physical' values of the resonance parameters, one would need both chiral and continuum extrapolations.
 It should be noted that in QCD, the coupling is known to have a mild chiral dependence.

\section{Conclusions}
        Composite Higgs models, which address the naturalness problem, can be studied using lattice gauge theory techniques. 
        This gives access to nonperturbative quantities from first principles. We focused on the scattering of two pseudoscalars in the vector resonance channel - it corresponds to $\pi\pi\to\rho\to\pi\pi$ scattering in QCD.
        If this theory was the underlying gauge theory completing the Higgs sector of the SM as proposed in \cite{Cacciapaglia:2014uja}, this particular channel would contribute to EW boson scattering.
        By studying the properties of the resonance - its mass and width, we provide useful input to constrain the parameter space of the model in question.

        In this work we use the lattice QCD techniques and present the first result for the phase shift in the SU(2) model with 2 fundamental flavours. We found $g_{\rho\pi\pi} = 11(2)$, which somewhat larger than SU(3) value of 6.
        For the time being this result should be treated as preliminary, because it does not include chiral or continuum extrapolations needed to extract the physical value of $g_{\rho\pi\pi}$. 
        This is still work in progress and will be included in the future iteration of this project.

\vspace{0.3cm}

\textbf{Acknowledgments} 

\vspace{0.1cm}

We thank C. Pica for allowing us to use the computer time on the Abacus cluster in Odense.    
T.J is supported by UK STFC grants ST/L000458/1 and ST/P000630/1, 
S.P. acknowledges support by Research Agency ARRS (research core funding No. P1-0035 and No. J1-8137) and DFG grant No. SFB/TRR 55.
The initial steps of this project  have been performed on the HPC facilities at the HPCC centre of the University of Plymouth.   
This work used the DiRAC Complexity system, operated by the University of Leicester IT Services, which forms part of the STFC DiRAC HPC Facility (www.dirac.ac.uk). 
This equipment is funded by BIS National E-Infrastructure capital grant ST/K000373/1 and  STFC DiRAC Operations grant ST/K0003259/1. DiRAC is part of the National E-Infrastructure.

\end{document}